\documentclass{pnastwo}

\usepackage{enumerate}
\usepackage{grffile}
\usepackage{cite}

\newcommand\sign{\mathop{\mathrm{sign}}}

\url{www.pnas.org/cgi/doi/10.1073/pnas.0709640104}
\copyrightyear{2008}
\issuedate{Issue Date}
\volume{Volume}
\issuenumber{Issue Number}

\begin{document}

\title{Irrelevance of Bell's Theorem for experiments involving correlations in space and time: a specific loophole-free computer-example}

\author{Hans De Raedt\affil{1}{
Zernike Institute for Advanced Materials,
University of Groningen, Nijenborgh 4, NL-9747 AG Groningen, The Netherlands
},
Kristel Michielsen\affil{2}{
Institute for Advanced Simulation, J\"ulich Supercomputing Centre,
Forschungszentrum J\"ulich, D-52425 J\"ulich,
RWTH Aachen University, D-52056 Aachen,
Germany
}, Karl Hess\affil{3}{ Center for Advanced Study, University of Illinois, Urbana, Illinois}
}

\contributor{Member submission to the Proceedings of the National Academy of Sciences
of the United States of America}

\maketitle

\begin{article}
\begin{abstract}
John Bell is generally credited to have accomplished the remarkable ``proof" that any theory of physics, which is both
Einstein-local and ``realistic" (counterfactually definite), results in a strong upper bound to the correlations that are
measured in space and time. He thus predicts that Einstein-Podolsky-Rosen experiments cannot violate Bell- type inequalities. We present a
counterexample to this claim, based on discrete-event computer simulations. Our model-results fully agree with the predictions
of quantum theory for Einstein-Podolsky-Rosen-Bohm experiments and are free of the detection- or a coincidence-loophole.
\end{abstract}


\section{Introduction}\label{sub1}

\dropcap{C}ounterfactual ``measurements" yield values that have been derived by means other
than direct observation or actual measurement, such as by calculation on the basis of a well-substantiated theory.
If one knows an equation that permits deriving reliably expected values
from a list of inputs to the physical system under investigation,
then one has ``counterfactual definiteness'' (CFD) in the knowledge of that system.
For an extensive discussion of counterfactuals see Ref.~\cite{PEAR09}.

The derivation of the Theorem of Bell and Bell's inequality necessitate the postulate of counterfactual definiteness~\cite{HESS15b}.
In the present paper we adopt the definition of CFD as given in our earlier paper~\cite{HESS15b}:
\begin{center} \medskip \framebox{
\parbox[t]{0.8\hsize}{
A counterfactually definite theory is described by a function (or functions) that map(s) tests
onto numbers.The variables of the function(s) argument must be chosen in a one to one correspondence to physical entities that
describe the test(s) and must be independent variables in the sense that they can be arbitrarily chosen from their respective
domains.
}}
\medskip \end{center}

In brief, CFD means that the output state of a system can,
in principle, be calculated using an explicit formula, e.g. $y = f(x)$
where $f(.)$ is a known (vector-valued) function of its argument $x$.
If $x$ denotes a vector of values then, according to the above definition,
the elements of this vector must be independent variables for the mathematical model to be CFD-compliant.

Although it is clear that CFD cannot be tested in a conventional laboratory experiment~\cite{PERE95},
a digital computer is nothing but a physical device that performs a kind of experiment (e.g. flipping bits),
albeit one that is under perfect control (we assume that the computer is operating flawless).
Therefore a digital computer may be used as a metaphor for carrying out ideal, perfect experiments.
In particular, it is trivial to perform computer experiments using functions that
satisfy the criterion of a CFD theory.
With all this in mind, in the present paper we only consider formulas $y = f(x)$
that can be implemented as an algorithm running on a digital computer.
Using the digital computer as a metaphor guarantees that we have a well-defined precise (in terms of bits)
representation of the concepts and algorithms (also in terms of bits) involved.

The purpose of the paper is to scrutinize the relation and relevance of CFD to models of Einstein-Podolsky-Rosen-Bohm (EPRB)
experiments. We demonstrate that there exist both CFD and non-CFD-compliant computer models for the EPRB experiments that
produce results in complete agreement with those of quantum theory. Because these computer models do not contain quantum
concepts, CFD does not distinguish classical from quantum physics for the case of EPRB experiments.

\section{Computational model}\label{INTRO}

In this section we describe a loophole-free implementation of Bohm's version~\cite{BOHM51}
of the Einstein-Podolsky-Rosen thought experiments~\cite{EPR35}.
This implementation simulates laboratory (EPRB) experiments with photons~\cite{ASPE82b,WEIH98}
but does not suffer from the practical limitations of real experiments: the computer experiments that
we report upon are ideal, perfect, loophole free experiments.
The computational model of the EPRB experiment is constructed such that it can reproduce, {\sl exactly},
the single particle averages and two-particle correlations of the singlet state~\cite{RAED06c,MICH14a}.

We begin by specifying the model of the observation stations which
are considered to be identical computational units
which operate according to a specific algorithm, see Fig.\ 1.
Input to a unit is the setting $a$ (representing the angle of the polarizer),
an angle $0\le \phi < 2\pi$ (representing the polarization of the photon),
and a number $0\le r<1$.
Output of a given unit is a binary variable $x=\pm1$ (representing the detection event at one
of the two detectors placed behind the polarizer),
and a time-related parameter $0\le t^*\le T$ (related to the recorded time-tag).
The model parameter $T$ is fixed and does not depend on the setting $a$.

Upon receiving the input $(a,\phi,r)$ the unit executes the following two steps~\cite{RAED06c}:

\begin{eqnarray}
&1.&\mathrm{compute\ \ \ }
c=\cos[2(a-\phi)]\;,\; s=\sin[2(a-\phi)],
\label{cfd0a}
\\
&2.&\mathrm{set\ \ \ }
x=\sign(c) \;,\; t^*=rTs^2
.
\label{cfd0b}
\end{eqnarray}
These two lines form the core of the computer algorithm. The simplicity of this algorithm is enticing, however, it contains
several key features. One is the creation of a time-related variable $t^*$ that has the interesting property of being a function
of both the local setting $a$ and the angle $\phi$. In contrast to Bell and
Clauser-Horn-Shimony-Holt, we are dealing therefore, with time related parameters that depend on the local setting of each
station. In addition, the model introduces randomness by a number $0 \leq r \le 1$, distributed uniformly.

It is important to notice that the variable $t^*$ in Eq.~(\ref{cfd0b})
may be imagined as being related to a ``pointer-position" of a clock
that symbolizes dynamic many-body interactions of
the photon with the constituent particles of the source and local measurement
equipment (polarizers etc.). All of these particles perform a (relativistic) many-body ``dance" that certainly may
depend on the local equipment orientation as well as on properties of the incoming photons. Because this many body ``dance"
has never been explored in actual EPRB equipments, we consider $t^*$ in the following only as a computer generated time-related
tag that is used in order to deal selectively with the results for $x$ after the whole computer experiment is done.

For every input event $(a,\phi,r)$, we know the values
of all outputs variables $x=x(a,\phi)$ and $t^*=t^*(a,\phi,r)$.
Therefore, the input-output relation of this unit, represented by the diagram of Fig.\ 1, satisfies the requirement of CFD.
We also use below the somewhat simpler notation
$x(a)=x(a,\phi)$ and $t^*(a)=t^*(a,\phi,r)$, keeping in mind that the $x$'s depend on $\phi$ and
the $t^*$'s depend on both $\phi$ and $r$.
Here and in the following, it is implicitly understood that for every instance
of new input, the values of the $\phi$'s and $r$'s are generated ``at random''.
The procedure for generating the $\phi$'s is specified in section ``Computer simulation results''.

The computational equivalent of the EPRB experiment~\cite{ASPE82b,WEIH98} is shown in Fig.\ 2.
We start by assuming that the source $\mathbf{S}$ and the observation stations $i=1,2$
are equipped with idealized, perfect clocks (not shown)
that have been synchronized before the source is being activated.
Each time the source $\mathbf{S}$ is activated, two photons are sent in opposite directions.
The source is activated at times $\tau_1,\ldots,\tau_N$ and denote the
minimum time interval between two emission events by $\delta\tau=\min_{n=1,\ldots,N-1} (\tau_{n+1}-\tau_n)$.

Each photon traveling to observation station $i=1,2$
carries its data in the form of an angle $\phi_i$ (representing the polarization)
and a pseudo-random number $0<r_i<1$.
The purpose of $r_i$ is to account, be it in a highly over-simplified manner,
for the influence of the many-body interactions of the incoming photon with the
constituent particles of the measurement equipment (polarized beam splitter, retarders etc.) located at observation station $i$,
resulting in a change of the time-of-flight from the source to the detector at observation station $i$~\cite{HESS15a,HESS15b}.

\begin{figure}[t]
\begin{center}
\includegraphics[width=0.12\textwidth]{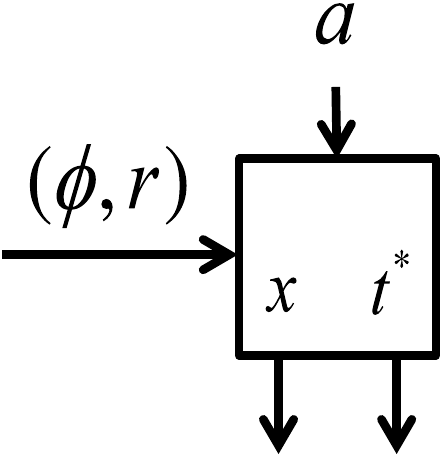}
\caption{%
Block diagram of an observation station. The input-output relations
$x=x(a,\phi)$ and $t^*=t^*(a,\phi,r)$ are defined by Eqs.~(\ref{cfd0a}) and (\ref{cfd0b}), respectively.
}
\end{center}
\label{cfdunit}
\end{figure}
Upon arrival of photon $n$ at station $i$, the observation station produces
the value $x_i=\pm1$ (see Eq.~(\ref{cfd0a})) and a time-tag

\begin{equation}
T_{i,n}=\tau_n + T_{\mathrm{TOF}}^{(i)} + t^*_{i,n}\;\quad,\quad i=1,2
.
\label{Tin}
\end{equation}
In laboratory EPRB experiments, one considers differences of time-tags only~\cite{WEIH98,HENS15}.
As the distances between the sources and the observation stations $i=1,2$ are fixed, we may assume that
the time $T_{\mathrm{TOF}}^{(i)}$ it takes the photon to reach the observation station $i$ is constant.
In general $T_{\mathrm{TOF}}^{(1)}\not=T_{\mathrm{TOF}}^{(2)}$ but the difference between the
two times-of-flight may be compensated for by adding this difference to the appropriate measured clock time $T_{i,n}$.
Hence, for simplicity 
we assume that $T_{\mathrm{TOF}}=T_{\mathrm{TOF}}^{(1)}=T_{\mathrm{TOF}}^{(2)}$.

From Eqs.~(\ref{cfd0b}) and (\ref{Tin}) it follows that $\tau_n + T_{\mathrm{TOF}} \le T_{i,n} \le \tau_n + T_{\mathrm{TOF}} + T$.
In the following, we only consider the case $T < \delta \tau$, meaning that the maximum delay time $T$
is smaller than the minimum time interval $\delta \tau$ between two emissions of a pair of photons.
The restriction $T < \delta\tau$ implies that $\tau_n + T_{\mathrm{TOF}} \le T_{i,n} < \tau_n + T_{\mathrm{TOF}} + \delta\tau$
and this inequality has an important consequence because it guarantees that
there is a one to one correspondence between the value of the time-tag $T_{i,n}$ and the number $n$ of the emission event.
Thus, in contrast to actual experimental data~\cite{WEIH98}, for the data generated by the computer model
there is a one to one correspondence between the value of the time-tag $T_{i,n}$ and the number $n$ of the emission event
if the condition $T < \delta \tau$ is satisfied.
With this condition we have $T_{1,n}-T_{2,n}=t^*_{1,n} - t^*_{2,n}$
and therefore, in the computer model, the time-of-flight and photon emission times $\tau_1,\ldots,\tau_N$
are superfluous and may be omitted.
There is no coincidence loophole and as every photon arriving at an observation station
also produces an output event $(x(a),t^*(a))$, there is no detection loophole either.
Note also that the assumption $T <\delta\tau$
prevents the incorrect inclusion of impossible events, such as two different polarizer-settings for the same
measurement (for details on this see~\cite{HESS15b}).

Upon arrival of the photon, observation station $i=1,2$ executes the algorithm defined by
Eqs.~(\ref{cfd0a}) and (\ref{cfd0b}) and produces output in the form of the pair $(x_i=\pm1, 0\le t^*_i\le T)$.
The algorithm represented by Fig.\ 2 computes the vector-valued function
\begin{eqnarray}
\left(\begin{array}{r}
        x_1 \\
        t^*_1\\
        x_2\\
        t^*_2
\end{array}\right)
&=&
\left(\begin{array}{c}
        x_1(\phi_1,a_1) \\
        t^*_1(\phi_1,a_1,r_1)\\
        x_2(\phi_2,a_2)\\
        t^*_2(\phi_2,a_2,r_2)
\end{array}\right)
\nonumber \\
&=&
F(\phi_1,\phi_2,r_1,r_2,a_1,a_2)
,
\label{cfd1}
\end{eqnarray}
hence, according to the definition of CFD, Eq.~(\ref{cfd1}) defines a CFD-compliant theoretical model.

\begin{figure}[t]
\begin{center}
\includegraphics[width=0.35\textwidth]{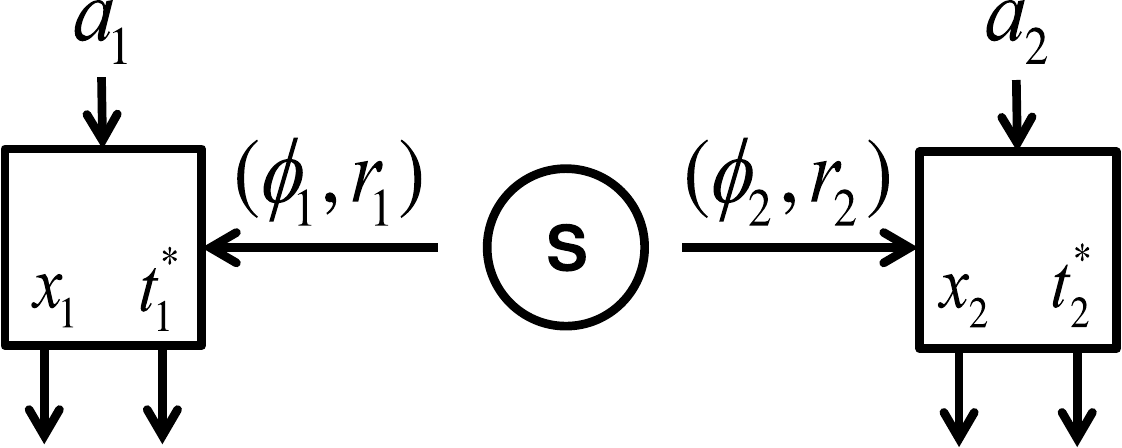}
\caption{Schematic layout of the computational equivalent of a laboratory EPRB experiment~\cite{ASPE82b,WEIH98}.
The input-output relation for $i=1,2$ is given by $x_i=x(a_i,\phi_i)$ and $t^*_i=t^*(a_i,\phi_i,r_i)$  where
$x=x(a,\phi)$ and $t^*=t^*(a,\phi,r)$ are defined by Eqs.~(\ref{cfd0a}) and (\ref{cfd0b}), respectively.
Alternatively, the input-output relation may be written as $(x_1,t^*_1,x_2,t^*_2)=F(\phi_1,\phi_2,r_1,r_2,a_1,a_2)$
showing that for fixed $(a_1,a_2)$ the simulation model satisfies the definition of a CFD theory~\cite{HESS15b}.
}
\end{center}
\label{fig0}
\end{figure}

The primary aim of EPRB experiments is to demonstrate a violation of the Bell-CHSH inequality under Einstein-local conditions~\cite{BELL01}.
By construction, the computer models that we use are metaphors for Einstein-local experiments:
changing $a_1$ ($a_2$) never has an effect on the values
of $x_2$ ($x_1$) or $t^*_2$ ($t^*_1$), not in the past nor in the future, hence
the output of one particular unit depends on the input to that particular unit only.

The crucial point that leads to a violation of Bell's theorem is now the following. We deal with photon pairs for which an
``entanglement" is defined in the Hilbert space of quantum mechanics. The correlations of the event of measurement of these
pairs are, on the other hand, obtained by measurements in ordinary space and time. In other words, some criterion
is employed to relate the measured pairs and identify them as belonging together. Such identification can be achieved, among
other possibilities, by use of two synchronized clocks indicating time $t$ in both measurement stations.
As soon as such identification and corresponding selection of pair-measurements is implemented,
we may derive a joint frequency distribution $P(T_1,T_2)$ for the time tags $T_{1,n}$ and $T_{2,n}$
for finding both $T_{1,n}$ and $T_{2_n}$
within a time-range $W$ around a time $t_n=\tau_n + T_{\mathrm{TOF}}$ of the synchronized station clocks.
This joint frequency distribution is derived in an Einstein local way and depends on the settings of the polarizers of {\bf
both} stations, a fact that cannot be accommodated in Bell-type proofs.

Bell-CHSH inequality tests require four different experiments with different choices of the settings of the observation stations.
Specifically, the setting of observation station $i=1,2$ can take two values which we denote by $(a_i,a_i')$.
The choice of setting $a_i$ or $a_i^\prime$ may be made at random~\cite{ASPE82b,WEIH98,HENS15}.
In real experiments, it takes a certain time to switch from one setting to another
but this time is less than the average time between two emission events~\cite{WEIH98}.
In the computer experiment, being an idealized perfect experiment, the algorithm is such that this cannot be an issue.

The algorithm represented by Fig.\ 2 is CFD-compliant.
However, the computational model represented by the diagram in Fig.\ 2
{\bf cannot} compute, e.g. $(x_1,x_2,x_1^\prime,x_2^\prime)$
with $x_i=x_i(a_i,\phi_i)$ and $x_i^\prime=x_i^\prime(a_i^\prime,\phi_i)$
in a CFD-compliant manner: it suffers from the so-called contextuality loophole~\cite{NIEU11}
because there is no guarantee that the random $\phi_i$'s used to compute the $x_i$'s
will be the same as the random $\phi_i$'s used for the calculation of the $x_i^\prime$'s.
Under these circumstances the correlations calculated from the data generated by the
model shown in Fig.\ 2 do not need to satisfy a Bell-type inequality~\cite{NIEU11,RAED11a,HESS15a,HESS15b} and,
as demonstrated explicitly below through simulation, indeed they do not.
Thus, the model of Fig.\ 2 {\bf cannot} be used to perform a
CFD-compliant simulation of the EPRB experiment.

The layout of a CFD-compliant computer model of the EPRB experiment is depicted in Fig.\ 3.
It uses the same units as the non-CFD-compliant model shown in Fig.\ 2, the only
difference being that the input $(\phi_i,r_i)$ is now fed into
an observation station with setting $a_i$ and another one with setting $a_i^\prime$.
As each of the four units operates according to the rules given by Eq.~(\ref{cfd0a}) and (\ref{cfd0b}), we have
$(x_1,x_1^\prime,x_2,x_2^\prime)=X(\phi_1,\phi_2,a_1,a_1^\prime,a_2,a_2^\prime)$
and
$(t^*_1,t^{*\prime}_1,t^*_2,t^{*\prime}_2)=T(\phi_1,\phi_2,r_1,r_2,a_1,a_1^\prime,a_2,a_2^\prime)$.
As the arguments of the functions $X$ and $T$ are independent and may take any
value out of their respective domain, the whole system represented by Fig.\ 3
satisfies, by construction, the criterion of a CFD theory.

\begin{figure*}[t]
\begin{center}
\includegraphics[width=0.8\hsize]{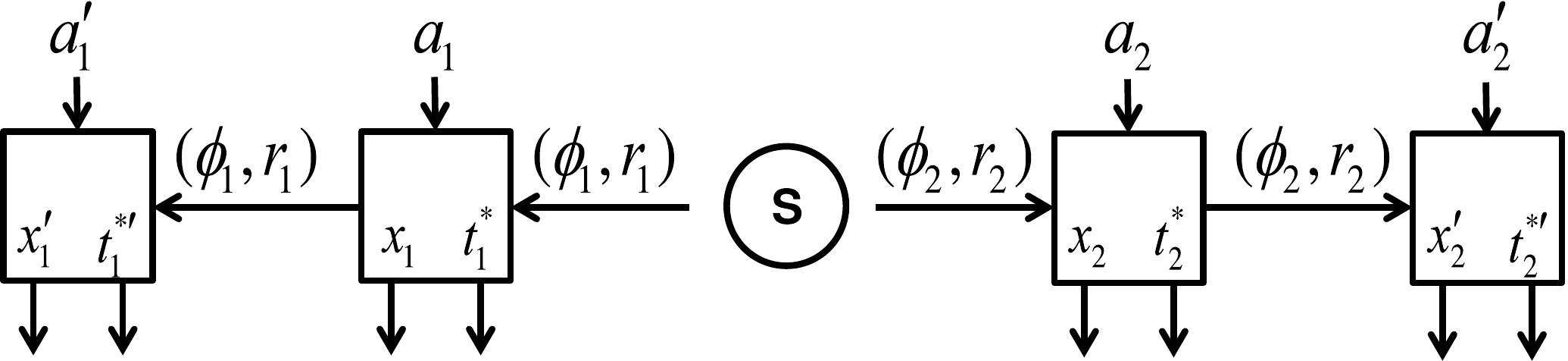}
\caption{Computational model for the EPRB experiment satisfying the criterion of a CFD theory.
}
\label{fig1}
\end{center}
\end{figure*}

\section{Bell-CHSH inequality and time-coincidence criterion}\label{sub2}

As CFD is used as at least one of the assumptions to prove the Bell-CHSH inequality~\cite{HESS15b},
it is instructive to see how this feature appears in the computational model.
Therefore, let us start by explicitly ignoring the $t$-variables.
As is clear from Fig.\ 3, the two stations
on the left of the source {\bf S} receive the same data $(\phi_1,r_1)$ from the source.
The settings $a_1$ and $a_{1}^\prime$ are fixed for the duration of the $N$ repetitions of the experiment.
The same holds for the two stations at the right of the source, with subscript 1 replaced by 2.
Clearly, each quadruple of output data $(x_1,x_1^\prime,x_2,x_2^\prime)$ is generated in a CFD-compliant manner.

For each input event (labeled by $n=1,\ldots,N$) we compute
\begin{eqnarray}
s_n&=& x_1 x_2 - x_1 x_2^\prime + x_1^\prime x_2 + x_1^\prime x_2^\prime
\nonumber \\
&=& x_1 \left(x_2 - x_2^\prime\right) + x_1^\prime \left( x_2 + x_2^\prime\right)
,
\label{cfd1a}
\end{eqnarray}
and
\begin{eqnarray}
S&=&\frac{1}{N}\sum_{n=1}^N s_n
.
\label{cfd1b}
\end{eqnarray}
where $N$ denotes the number of pairs that was processed by the units.
From Eq.~(\ref{cfd1a}) it follows immediately that $|s_n|\le 2$ and hence $|S|\le 2$.
Of course, this is what we expect: if the whole system is CFD-compliant and we ignore the $t$-variables,
we generate quadruples and then it is impossible to violate the Bell-CHSH inequality $|S|\le2$~\cite{BO1862,RAED11a}.

Next, we examine what happens if the time-tag variables $T_{i,n}$ are included.
In real EPRB experiments with photons, it is essential to use time-coincidence to identify pairs~\cite{ASPE82b,WEIH98,HENS15}.
The standard procedure adopted in these experiments is to introduce a time window $W$ and reject pairs that do not
satisfy the condition $|T_{1,n}-T_{2,n}|\le W$ (and similar for other relevant combinations of $T$'s)~\cite{WEIH98}.
The computational model defined by Eqs.~(\ref{cfd0a}) and (\ref{cfd0b})
together with the time-coincidence criterion yields, in the limit
that the time-window $W$ vanishes, the correlation of the singlet state~\cite{RAED06c,MICH14a} if we repeat
the experiment pair-wise, i.e. with four pairs of settings (see Fig.\ 2),
in which case the CFD criterion is clearly not satisfied.
We emphasize that unlike in the laboratory experiment,
in the computer experiment all pairs are created ``on demand'', each pair is detected, and
the time window only serves as a vehicle to post select pairs, not to identify them.
Post-selection only serves to ``probe'' the complicated, time-dependent many-body physics
that is involved when the photon passes through the optical system and triggers the detector.
In this sense, the computer experiment suffers from none of the loopholes that may occur in experiments.

Although not feasible with photons, using the computer as a metaphor we can perform
the ideal, loophole-free experiment satisfying all the requirements of a CFD theory.
In the remainder of this section
we discuss the ramifications to the Bell-type inequality that ensue when time is included in the description.
In the next section, we demonstrate that the CFD-compliant model reproduces the
quantum theoretical results of the singlet state.

We formalize the effect of the time-coincidence window by introducing the binary variables
\begin{eqnarray}
w_(a_1,a_2)&=&\Theta(W-|T_{1,n}-T_{2,n}|)
\nonumber \\
w_(a_1,a_2^\prime)&=&\Theta(W-|T_{1,n}-T^{\prime}_{2,n}|)
\nonumber \\
w_(a_1^\prime,a_2)&=&\Theta(W-|T^{\prime}_{1,n}-T_{2,n}|)
\nonumber \\
w_(a_1^\prime,a_2^\prime)&=&\Theta(W-|T^{\prime}_{1,n}-T^{\prime}_{2,n}|)
,
\label{cfd2}
\end{eqnarray}
where $\Theta(x)$ is the unit step function.
In essence, we extend the computational device by taking the output
of the two units described earlier and feeding the time-tag output
in a ``correlator'' that computes, for each event $n$, the four
binary variables defined by Eq.~(\ref{cfd2}).
Adding the correlator does not change the fact that the computer model is CFD-compliant.
Indeed, a given input $(\phi_1,r_1,\phi_2,r_2)$ together with the settings $(a_1,a_1^\prime,a_2,a_2^\prime)$
completely determines the values of all (two-valued) output variables
$x(a_1)$, $x(a_1^\prime)$, $x(a_2)$, $x(a_2^\prime)$, $w(a_1,a_2)$, $w(a_1,a_2^\prime)$, $w(a_1^\prime,a_2)$, and $w(a_1^\prime,a_2^\prime)$.
Note that e.g. $w(a_1,a_2)=0$ means that the particular pair has been discarded
by the time-coincidence criterion for the pair of settings $(a_1,a_2)$ but
that this does not imply that e.g. $w(a_1,a_2^\prime)=0$.
In other words, the values of the $w$'s are used to post-select pairs.

Next we compute averages and correlations of the coincident pairs as is done in laboratory EPRB experiments~\cite{WEIH98}.
The single-$x$ averages and correlation for the settings $(a_1,a_2)$ are defined by
\begin{eqnarray}
E_1(a_1,a_2)&=&\frac{\sum w(a_1,a_2)x(a_1)}{\sum w(a_1,a_2)}
\nonumber \\
E_2(a_1,a_2)&=&\frac{\sum w(a_1,a_2)x(a_2)}{\sum w(a_1,a_2)}
\nonumber \\
E(a_1,a_2)&=&\frac{\sum w(a_1,a_2)x(a_1) x(a_2)}{\sum w(a_1,a_2)}
,
\label{cfd3}
\end{eqnarray}
and we have similar expressions for the other choices of settings.
In Eq.~(\ref{cfd3}) 
it is understood that
$\sum$ means $\sum_{n=1}^N$, i.e. the sum over all input events,
characterized by values of the $r$'s and $\phi$'s.
It is not difficult to see that $E_1(a_1,a_2)$, $E_2(a_1,a_2)$ etc. are zero,
up to fluctuations. The reason is that $\phi \rightarrow \phi+\pi/2$
changes the sign of the $x$'s but has no effect on the values of the $t^*$'s (see Eq.~(\ref{cfd0b})).
Therefore, if the $\phi's$ uniformly cover $[0,2\pi[$,
the number of times $x=+1$ and $x=-1$ appear is about the same.

The usual strategy of deriving a Bell-like inequality for
$\widehat S= E(a_1,a_2)-E(a_1,a_2^\prime)+E(a_1^\prime,a_2)+E(a_1^\prime,a_2^\prime)$
does not work simply because not all $w$'s need to be one for the same event $n$~\cite{HESS15a}
but we can, without using probability theory, derive another inequality by following the strategy of Larsson and Gill~\cite{LARS04}.
Denoting the number of input events for which the four pairs of
setting simultaneously satisfy the coincidence criterion by $N^\prime$
and the maximum number of pairs per setting that
satisfies the coincidence criterion by $N_{\mathrm{max}}$,
we have $0\le \delta\equiv N'/N_{\mathrm{max}}\le 1$ and
it is straightforward to show (by repeated application of the triangle inequality) that
the following statement holds:
in the case that the time-coincidence criterion is used to post-select pairs,
the correlations cannot violate the inequality
\begin{equation}
\left|
E(a_1,a_2)-E(a_1,a_2^\prime)+E(a_1^\prime,a_2)+E(a_1^\prime,a_2^\prime)
\right|\le 4-2\delta
\
.
\label{cfd11}
\end{equation}
Therefore, if the algorithm generates all the variables
strictly in accordance with the criterion of a CFD theory, using the time-coincidence window
to post-select pairs does not lead to the Bell-CHSH inequality unless $\delta=1$ in which case
all $w$'s are equal to one and none of the pairs are discarded by the post-selection procedure.
The term $2\delta$ in Eq.~(\ref{cfd11}) is a measure for the number of pairs that have been
post-selected relative to the number of emitted pairs.

At this point, it is important to mention that in establishing Eq.~(\ref{cfd11}),
the specific computational model that we have used as a concrete realization is not essential:
as long as the algorithm generates $x$'s and $T$'s in accordance with
the criterion of a CFD theory and $\delta>0$, Eq.~(\ref{cfd11}) holds.

\section{Computer simulation results}\label{OVER}

As explained earlier, from the logical structure of the algorithm it is evident that
the outcome of a particular unit cannot be influenced
by the input/output of another unit, not by the current input event nor by past or future events.
Therefore, all models that we consider generate data by a process that complies with the notion of Einstein locality.

We present the results of four different modes of simulating the EPRB experiments.
This section reports the results of simulations
for 100 repetitions of the EPRB experiment with $N=10^6$ events
per pair of settings in the case of the non-CFD-compliant models and
$N=4\times 10^6$ events for the CFD-compliant models.
We set $\phi_1=\phi$ and $\phi_2=\phi+\pi/2$ where $0\le \phi < 2\pi$ is a uniform pseudo-random number,
corresponding to the case in which the polarizations of the two photons within a pair are orthogonal and fully correlated
(if $\phi_1$ and $\phi_2$ are uncorrelated and random,
the average of the $x$'s and the average of e.g. $x_1 x_2$ are all zero, independent of the settings).
The algorithm of the unit simulating the observation station is defined by Eqs.~(\ref{cfd0a}) and~(\ref{cfd0b}).
For the settings we take $a_1=0$, $a_1^\prime=\pi/4$,
$a_2=\pi/8$, $a_2^\prime=3\pi/8$ as these are known to be a choice
that maximizes $S$, for the time window we take $W=1$ and
the maximum time delay is taken to be $T=1000$.

\begin{description}
\item[Case 1:] {\bf non-CFD-compliant model (see Fig.\ 2), no post selection by a time window.}
The averages and correlations are obtained
from four sets of measurements with the four different pairs of settings
$(a_1,a_2)$, $(a_1,a_2^\prime)$, $(a_1^\prime,a_2)$, and $(a_1^\prime,a_2^\prime)$.
As the output of the stations with say setting $(a_1,a_2)$ is not
available when the experiment runs with another setting, say $(a_1,a_2^\prime)$,
this computer experiment does not satisfy the criterion of a CFD theory,
nor does it mimic a real EPRB experiment with photons.
For a set of 100 repetitions, the simulations show that
54 out of 100 repetitions yield a violation of $|S|\le2$.
The average of $S$ being $2.0000$ with standard deviation $0.0016$.
Therefore, in a mathematically strict sense, for finite $N$,
the non-CFD-compliant model without post selection by a time window
yields data that violates the inequality $|S|\le2$, as expected~\cite{HESS15a}.
However, this model does not yield the correlation that
is characteristic for a quantum system in the singlet state.

\medskip
\item[Case 2:] {\bf non-CFD-compliant model (see Fig.\ 2), post selection by a time window.}
The averages and correlations are obtained
from four sets of measurements with four different pairs of settings
$(a_1,a_2)$, $(a_1,a_2^\prime)$, $(a_1^\prime,a_2)$, and $(a_1^\prime,a_2^\prime)$.
This computer experiment does not satisfy the criterion of a CFD theory.
Models that incorporate post selection 
are known to produce results that violate $|S|\le2$~\cite{PEAR70,FINE74,PASC86,BRAN87,RAED06c,RAED07b,ZHAO08,KHRE15b}.
For a set of 100 repetitions, the simulations show
that 100 out of 100 repetitions yield a violation of $|S|\le2$,
the average of $S$ being $2.824$ with standard deviation $0.032$.
This value of $S$ is very close to the theoretical maximum $2\sqrt{2}=2.8284$
for the quantum system in the singlet state~\cite{CIRE80}.

\medskip
\item[Case 3:] {\bf CFD-compliant model (see Fig.\ 3), no post selection by a time window.}
If the averages and correlations are obtained
from sets of quadruples $(x_1,x_1^\prime,x_2,x_2^\prime)$,
the model is CFD-compliant and the CHSH inequality $|S|\le2$ cannot be violated,
because no relation to space-time variables is established.
For the same set of 100 repetitions as used in Case 1, the simulations show that
0 out of 100 repetitions yield $|S|>2$, the average of $S$ being $2$ with standard deviation $0$,
hence the CHSH inequality is satisfied.

\medskip
\item[Case 4:] {\bf CFD-compliant model (see Fig.\ 3), post selection by a time window.}
If the averages and correlations are obtained
from sets of quadruples $(x_1,x_1^\prime,x_2,x_2^\prime)$ post-selected
by way of the time-coincidence criterion,
the CHSH inequality cannot be derived
and the correlations satisfy instead Eq.~(\ref{cfd11}).
For the same set of 100 repetitions as used in Case 1, the simulations show that
100 out of 100 repetitions yield a violation of $|S|\le2$,
the average of $S$ being $2.827$ with standard deviation $0.016$.
This value of $S$ is very close to the theoretical maximum $2\sqrt{2}=2.8284$
for the quantum system in the singlet state~\cite{CIRE80}.
The minimum value of $\delta$ found in these 100 repetitions
is $0.14\times 10^{-3}$, that is the number of pairs rejected by
the time-coincidence criterion is significant.
Note that unlike case 2, it is impossible to perform
this CFD-compliant experiment with photons.

\medskip
In Fig.\ 4 we show the correlation $E(a_1,a_2)$ and single-$x$ averages as obtained
from the simulation with the CFD-compliant model with post selection.
Quantum theory predicts $E(a_1,a_2)=-\cos[2(a_1-a_2)]$.
From Fig.\ 4 it is clear that the CFD-compliant model reproduces the results of quantum theory
without making any reference to the latter.

\end{description}

\begin{figure}[t]
\begin{center}
\includegraphics[width=0.5\textwidth]{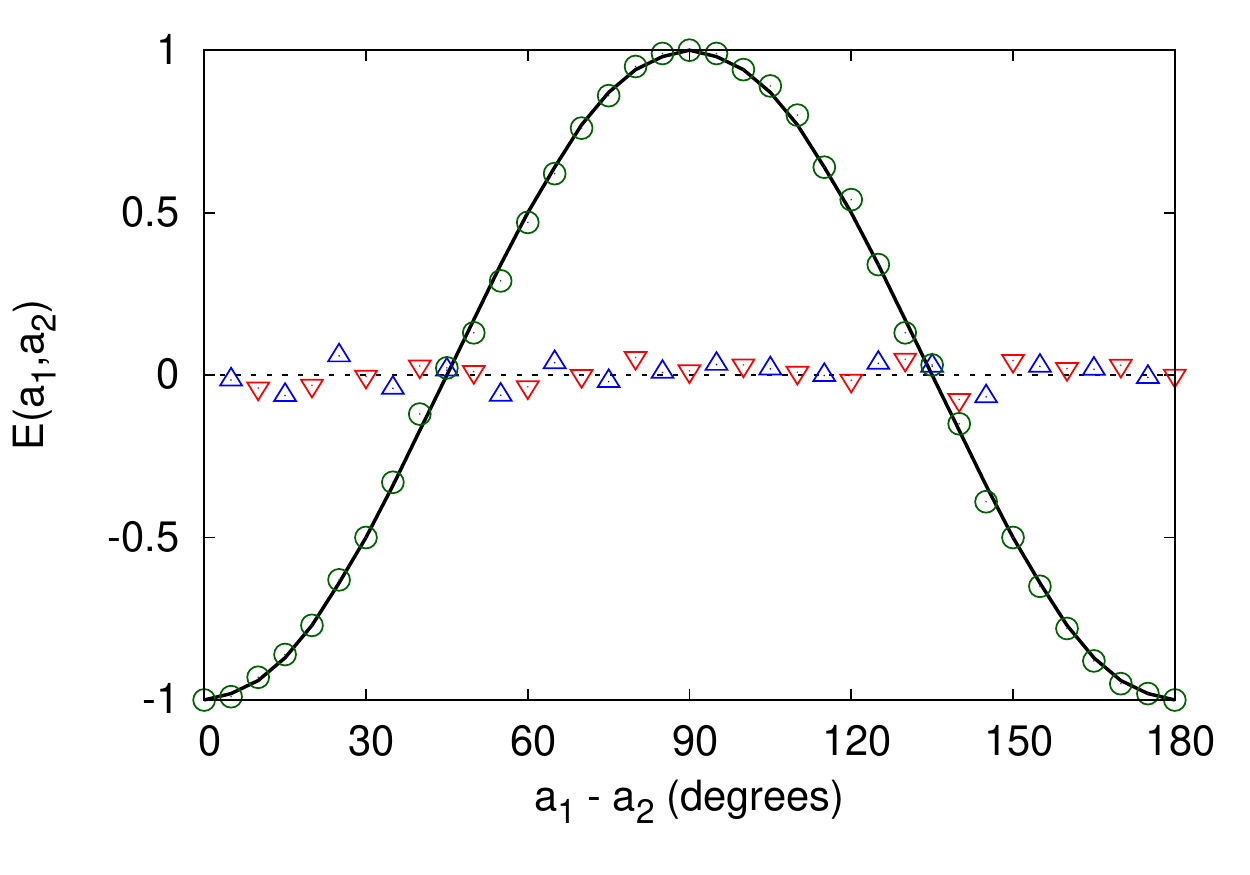}
\caption{%
The correlation $E(a_1,a_2)$ ($\bigcirc$) and single-$x$ averages $E_1(a_1,a_2)$ ($\bigtriangleup$)
and $E_2(a_1,a_2)$ ($\bigtriangledown$)
as a function of $a_1-a_2$ as obtained from a computer simulation data of the CFD-compliant model (see Fig.\ 3)
with time-coincidence window $W=1$ and $T=1000$.
Solid line: quantum theoretical result of the correlation $E(a_1,a_2)$
of a quantum system in the singlet state.
Dashed line: quantum theoretical result of the single-$x$ averages $E_1(a_1,a_2)=E_2(a_1,a_2)=0$ in the singlet state.
}%
\label{fig2}
\end{center}
\end{figure}

\section{Conclusion}
A CFD-compliant model of the EPRB experiment that incorporates post-selection by a time window
can violate the inequality $|S|\le2$ but cannot violate Eq.~(\ref{cfd11}).
Furthermore, with the proper choice of model parameters, this model reproduces
the results of the quantum theoretical description of the EPRB experiment in terms of the singlet state.
Therefore, we have demonstrated that in the case of the EPRB experiment,
CFD does not separate nor distinguish classical from quantum physics.
The CFD-compliant model, which may be viewed as having physical time involved in the post-selection process as a hidden variable,
provides a counter example to the dogma that CFD implies a Bell-type inequality.



\begin{thebibliography}{10}

\bibitem{PEAR09}
Pearl, J.
\newblock (2000) {\em {Causality: models, reasoning, and inference}}.
\newblock (Cambridge University Press, Cambridge).

\bibitem{HESS15b}
Hess, K, {De Raedt}, H,  \& Michielsen, K.
\newblock (2015) Counterfactual definiteness and {Bell's} inequality.
\newblock {\em Submitted to PNAS} {\bf ??}, ?? -- ??

\bibitem{PERE95}
Peres, A.
\newblock (1995) {\em Quantum Theory: Concepts and Methods}.
\newblock (Kluwer Academic Publishers, Dordrecht, Boston London), pp. 2006 --
  2007.

\bibitem{BOHM51}
Bohm, D.
\newblock (1951) {\em Quantum Theory}.
\newblock (Prentice-Hall, New York).

\bibitem{EPR35}
Einstein, A, Podolsky, A,  \& Rosen, N.
\newblock (1935) Can quantum-mechanical description of physical reality be
  considered complete?
\newblock {\em Phys. Rev.} {\bf 47}, 777 -- 780.

\bibitem{ASPE82b}
Aspect, A, Dalibard, J,  \& Roger, G.
\newblock (1982) Experimental test of {Bell}'s inequalities using time-varying
  analyzers.
\newblock {\em Phys. Rev. Lett.} {\bf 49}, 1804 -- 1807.

\bibitem{WEIH98}
Weihs, G, Jennewein, T, Simon, C, Weinfurther, H,  \& Zeilinger, A.
\newblock (1998) {Violation of {Bell}'s inequality under strict {Einstein}
  locality conditions}.
\newblock {\em Phys. Rev. Lett.} {\bf 81}, 5039 -- 5043.

\bibitem{RAED06c}
{De Raedt}, K, Keimpema, K, {De Raedt}, H, Michielsen, K,  \& Miyashita, S.
\newblock (2006) {A local realist model for correlations of the singlet state}.
\newblock {\em Eur. Phys. J. B} {\bf 53}, 139 -- 142.

\bibitem{MICH14a}
Michielsen, K \& {De Raedt}, H.
\newblock (2014) {Event-based simulation of quantum physics experiments}.
\newblock {\em Int. J. Mod. Phys. C} {\bf 25}, 01430003.

\bibitem{HESS15a}
Hess, K, {De Raedt}, H,  \& Michielsen, K.
\newblock (2015) From {Boole} to {Leggett-Garg}: Epistemology of {Bell}-type
  inequalities.
\newblock {\em Submitted to PNAS} {\bf ??}, ?? -- ??

\bibitem{HENS15}
Hensen, B, Bernien, H, Dreau, A.~E, Reiserer, A, Kalb, N, Blok, M.~S,
  Ruitenberg, J, Vermeulen, R. F.~L, Schouten, R.~N, Abellan, C, Amaya, W,
  Pruneri, V, Mitchell, M.~W, Markham, M, Twitchen, D.~J, Elkouss, D, Wehner,
  S, Taminiau, T.~H,  \& Hanson, R.
\newblock (2015) Loophole-free {Bell} inequality violation using electron spins
  separated by 1.3 kilometres.
\newblock {\em Nature} p. 15759.

\bibitem{BELL01}
Bell, J.~S.
\newblock (2001) {\em On the Foundations of Quantum Mechanics}.
\newblock (World Scientific, Singapore, New Jersey, London, Hong Kong), pp. 228
  -- 229.

\bibitem{NIEU11}
Nieuwenhuizen, T.~M.
\newblock (2011) Is the contextuality loophole fatal for the derivation of
  {Bell} inequalities?
\newblock {\em Found. Phys.} {\bf 41}, 580 -- 591.

\bibitem{RAED11a}
{De Raedt}, H, {Hess}, K,  \& {Michielsen}, K.
\newblock (2011) {Extended Boole-Bell inequalities applicable to quantum
  theory}.
\newblock {\em J. Comput. Theor. Nanosci.} {\bf 8}, 1011 -- 1039.

\bibitem{BO1862}
Boole, G.
\newblock (1862) On the theory of probabilities.
\newblock {\em Phil. Trans. R. Soc. Lond.} {\bf 152}, 225 -- 252.

\bibitem{LARS04}
Larsson, J.-{\AA} \& Gill, R.~D.
\newblock (2004) {Bell}'s inequality and the coincidence-time loophole.
\newblock {\em Europhys. Lett.} {\bf 67}, 707 -- 713.

\bibitem{PEAR70}
Pearle, P.~M.
\newblock (1970) Hidden-variable example based upon data rejection.
\newblock {\em Phys. Rev. D} {\bf 2}, 1418 -- 1425.

\bibitem{FINE74}
Fine, A.
\newblock (1974) {On the completeness of quantum theory}.
\newblock {\em Synthese} {\bf 29}, 257 -- 289.

\bibitem{PASC86}
Pascazio, S.
\newblock (1986) Time and {Bell}-type inequalities.
\newblock {\em Phys. Lett. A} {\bf 118}, 47 -- 53.

\bibitem{BRAN87}
Brans, C.
\newblock (1987) {Bell}'s theorem does not eliminate fully causal hidden
  variables.
\newblock {\em Int. J. Theor. Phys.} {\bf 27}, 219 -- 226.

\bibitem{RAED07b}
{De Raedt}, K, {De Raedt}, H,  \& Michielsen, K.
\newblock (2007) {A computer program to simulate Einstein-Podolsky-Rosen-Bohm
  experiments with photons}.
\newblock {\em Comp. Phys. Comm.} {\bf 176}, 642 -- 651.

\bibitem{ZHAO08}
{Zhao}, S, {De Raedt}, H,  \& Michielsen, K.
\newblock (2008) {Event-by-event simulation model of
  Einstein-Podolsky-Rosen-Bohm experiments}.
\newblock {\em Found. Phys.} {\bf 38}, 322 -- 347.

\bibitem{KHRE15b}
Khrennikov, A.
\newblock (2015) {CHSH} inequality: Quantum probabilities as classical
  conditional probabilities.
\newblock {\em Found. Phys.} {\bf 45}, 711 -- 725.

\bibitem{CIRE80}
Cirel'son, B.~S.
\newblock (1980) Quantum generalizations of {Bell}'s inequality.
\newblock {\em Lett. Math. Phys.} {\bf 4}, 93 -- 100.

\end{thebibliography}

\end{article}
\end{document}